\documentclass[11pt]{article}

\begin{document}
\def\a{\alpha}\def\b{\beta}\def\g{\gamma}\def\d{\delta}\def\e{\epsilon }
\def\k{\kappa}\def\l{\lambda}\def\L{\Lambda}\def\s{\sigma}\def\S{\Sigma}
\def\Th{\Theta}\def\th{\theta}\def\om{\omega}\def\Om{\Omega}\def\G{\Gamma}
\def\y{\vartheta}\def\m{\mu}\def\n{\nu}
\def\ws{worldsheet}
\def\susy{supersymmetry}
\def\ts{target superspace}
\def\ks{$\k$--symmetry}
\newcommand{\plabel}{\label}
\renewcommand\baselinestretch{1.5}
\newcommand{\nn}{\nonumber\\}\newcommand{\p}[1]{(\ref{#1})}
\renewcommand{\thefootnote}{\fnsymbol{footnote}}
\thispagestyle{empty}
\begin{flushright}
hep--th/0110192
\end{flushright}


\bigskip
\thispagestyle{empty}

\vspace{1cm}
\begin{center}
{\Large\bf On the $M5$ and the $AdS_7/CFT_6$ Correspondence}

\vspace{2.5cm}
A. J. Nurmagambetov \footnote{Also at Institute for Theoretical
Physics NSC KIPT, Kharkov, Ukraine} \footnote{Electronic address:
ajn@rainbow.physics.tamu.edu} and I. Y. Park \footnote{Electronic
address: ipark@rainbow.physics.tamu.edu}

\vspace{1.8cm}
{\small\it Center for Theoretical Physics\\
Texas A\&M University\\
College Station, 77843 TX, USA}

\vspace{3.3cm} {\bf Abstract}
\end{center}
The chiral primary operators of the $D=6$ superconformal $(2,0)$ theory
corresponding to 14 scalars of N=4 D=7 supergravity are obtained by
expanding the world volume action for the M5-brane around an $AdS_7\times S^4$
background. In the leading order, the operators take their values in the symmetric traceless
representation of the $SO(5)$ R-symmetry group in consistency with the early
conjecture on their structure based on the superconformal symmetry and Matrix-like model arguments.

\vspace{3cm} {\it PACS: 11.15.-q; 11.17.+y}

\renewcommand{\thefootnote}{\arabic{footnote}}
\setcounter{page}1

\newpage
\section{Introduction}

The AdS/CFT correspondence \cite{jm}, \cite{gkp}, \cite{witten}
has revived the interest in superconformal theories and
$AdS_{D-p}\times S^p$ configurations in supergravity (SUGRA)
theories. During the past several years the correspondence between
supergravity modes and super-Yang-Mills (SYM) operators was
verified through different methods. (See Ref. \cite{agmoo} for
an extensive list of references.) Many results have been obtained
for $AdS_5/CFT_4$ correspondence. In the cases of $AdS_4/CFT_3$
and $AdS_7/CFT_6$ a relatively smaller number of papers were
written.

The $AdS_4$ and $AdS_7$ geometry arise \cite{gt} in the large $N$
limit of $N$-coincident M2-branes \cite{bst}, \cite{bst1} and
M5-branes \cite{pst}, \cite{blnpst}, \cite{apps} respectively.
\footnote{Covariant equations of motion for the M5-brane were
obtained in \cite{sezgin} from the superembedding approach (see,
e.g., \cite{bpstv}, and \cite{embed} for recent reviews).
Relations between different formulations were established in
\cite{blnpst1}.} In this paper we focus on the case of the
$AdS_7/CFT_6$ correspondence. The structure of the CFT operators
was obtained by analyzing the representations of superalgebra
$Osp(8^*|4)$ \cite{minwalla}, \cite{aoy}, \cite{lr}. Then the
correspondence between CFT operators and supergravity modes can be
established by comparing the various quantum numbers of their
representations of $Osp(8^*|4)$.  Based on such considerations it
was conjectured in \cite{minwalla1} that the chiral primaries of
the (2,0) CFT are scalar operators in the symmetric traceless
reps. of the R-symmetry group $SO(5)$. This conjecture is in
accordance with results \cite{seiberg}, \cite{abs} obtained from
the Matrix-like DLCQ description of six-dimensional $(2,0)$
superconformal field theory as a quantum mechanics on the moduli
space of instantons.

Another way of matching various boundary CFT operators with the
supergravity modes was proposed in \cite{das}. The authors derived
the SYM operators that are dual to the longitudinally polarized
NS-NS two-form gauge field by expanding D3-brane action around an
$AdS_5\times S^5$ background.\footnote{Further related discussions
can be found in \cite{iyp,dt2,rr}.} In \cite{tran} this approach
was applied to obtain the CFT operators that correspond to 20
scalar modes of the five dimensional gauged supergravity. For the
computation the Kaluza-Klein reduction \cite{nieuw0,dnp} ansatz
obtained in \cite{clpst} was used.

Here following \cite{das} and \cite{tran}, we will consider the
expansion of the M5-brane action around $AdS_7\times S^4$
background.  In Section 2 we briefly review the structure of the
non-linear ansatz for reduction of eleven-dimensional supergravity
on $S^4$ and the equations of motions of seven-dimensional
supergravity \cite{nieuw,pope}. For our aim, consideration of the
bosonic subsectors is sufficient. To simplify the calculations, in
Section 3 we set seven-dimensional gauge fields to zero, which
imposes additional constraints on the scalar sector of the D=7
supergravity. This, in turn, allows us to choose a diagonal
parameterization for the scalar matrix. After substituting the
ans\"atze for metric and target space gauge fields into the M5
action, we work out the CFT operators expanding the M5 action to
linear order in the diagonal modes. Finally, we relax the zero
setting of the gauge fields and obtain the CFT operators
corresponding to the full set of the scalar fields. The last
section contains our conclusions.

\section{D=11 and D=7 SUGRA analysis}

The starting point is the action of D=11 SURGA \cite{CJS}
$$
S_{CJS}=\int \, d^{11}x \sqrt{-{\hat g}} [{\hat R}({\hat
\omega})+\dots ]
$$
\begin{equation}\plabel{cjs}
-\int
\, d^{11}x \sqrt{-{\hat g}}{1\over 2!4!}{\hat
F}^{(4)}_{\underline{\hat m}_1\dots
\underline{\hat m}_4}{\hat F}^{(4)\underline{\hat m}_1\dots
\underline{\hat m}_4}
-\int_{{\cal M}^{11}} \, {1\over 6}
{\hat A}^{(3)}\wedge {\hat F}^{(4)}\wedge {\hat F}^{(4)},
\end{equation}
where the ellipses denotes the terms involving the Rarita-Schwinger field.
The equation of motion for $A^{(3)}$ is
\begin{equation}\plabel{emA3}
d(\hat{\ast}{\hat F}^{(4)}-{1\over 2}{\hat A}^{(3)}\wedge {\hat
F}^{(4)})=0,
\end{equation}
which can be viewed as the first order Bianchi identity for the dual field strength \cite{bbs},
\cite{cjlp}
\begin{equation}\plabel{BI}
d{\hat F}^{(7)}=0;\qquad {\hat F}^{(7)}=d{\hat A}^{(6)}+{1\over 2}{\hat A}^{(3)}\wedge
{\hat F}^{(4)}.
\end{equation}

The equations of motion for N=4 D=7 $SO(5)$ gauged SUGRA
\cite{ppvn} can be obtained from the D=11 SUGRA by the use of
non-linear Kaluza-Klein $S^4$ reduction ansatz presented in
\cite{nieuw}, \cite{pope}. In the notation of \cite{popelec}, it
is given by
\begin{equation}\plabel{nlansds}
d{\hat s}^2_{11}=\tilde{\triangle}^{1/3}ds^2_7+g^{-2}
\tilde{\triangle}^{-2/3}T_{ab}^{-1}D{\mu}^a D{\mu}^b,
\end{equation}
\begin{equation}\plabel{nlansF4}
{\hat F}^{(4)}={1\over 4!}\e_{a_1\dots a_5} [-{1\over g^3}U\tilde{\triangle}^{-2}
\mu^{a_1}D\mu^{a_2}\wedge\dots\wedge D\mu^{a_5}
+{4\over g^3}\tilde{\triangle}^{-2}T^{a_1}_{~b}DT^{a_2}_{~c}\mu^b \mu^{c}
D\mu^{a_3}\wedge\dots \wedge D\mu^{a_5}
\end{equation}
$$
+{6\over g^2}\tilde{\triangle}^{-1} F^{(2)~a_1 a_2}\wedge D\mu^{a_3}\wedge D\mu^{a_4}
T^{a_5}_{~b}\mu^{b}]
-T_{ab}\ast C^{(3)a}\mu^b+{1\over g} C^{(3)}_a D\mu^a,
$$
\begin{equation}\plabel{nlansF7}
{\hat F}^{(7)}=-gU\epsilon_{(7)}-g^{-1}(T_{ab}^{-1}\ast DT_{bc})\wedge
(\mu^c D\mu^a)
\end{equation}
$$
+{1\over 2}g^{-2}T^{-1}_{ac}T^{-1}_{bd}\ast F^{(2)~ab}\wedge
D\mu^c\wedge D\mu^d+g^{-4}\tilde{\triangle}^{-1}T_{ab}~C^{(3)~a}
\mu^b \wedge W
$$
$$
-{1\over 6}g^{-3}\tilde{\triangle}^{-1}\e_{abcde}\ast C^{(3)~f} T^a_{~f}
T^b_{~g}\mu^g\wedge D{\mu}^c \wedge D{\mu}^d \wedge D{\mu}^e.
$$
Here
$$
U\equiv 2T_{ab}T_{bc}\mu^a\mu^c-\tilde{\triangle}T_{aa},\qquad
\tilde{\triangle}\equiv T_{ab}\mu^a\mu^b, \qquad
W={1\over 4!}\e_{a_1\dots a_5}\mu^{a_1} D{\mu}^{a_2}\wedge \dots \wedge
D{\mu}^{a_5},
$$
$$
F^{(2)}_{ab}=dA^{(1)}_{ab}+g~A^{(1)}_{ac}\wedge A^{(1)c}_{~~~~b},
$$
$$
DT_{ab}=dT_{ab}+gA^{(1)~c}_{a}T_{cb}+gA^{(1)~c}_{b}T_{ac},
$$
\begin{equation}\plabel{defs}
\mu^a\mu^a=1,\qquad D\mu^a=d\mu^a+g A^{(1)}_{ab}\mu^b,
\end{equation}
where $A^{(1)}_{ab}$ are the 10 gauge fields of $N=4$ $D=7$ gauged
supergravity. In \p{nlansds} -- \p{defs} $\epsilon_{(7)}$ is the
volume form on the seven-dimensional space-time and $T_{ab}$ is
a symmetric unimodular matrix of scalars in the 14'
representation of $SO(5)$ which admits the following
representation
\begin{equation}\plabel{Treps}
T_{ab}=(e^{S})_{ab},\qquad Tr~ S_{ab}=0.
\end{equation}

Substitution of the ansatz for ${\hat F}^{(4)}$ and ${\hat
F}^{(7)}=\hat{\ast}\hat F^{(4)}$ into the Bianchi identity for
${\hat F}^{(4)}$ and D=11 equation of motion \p{emA3} leads to the
following D=7 equations of motion
\begin{equation}\plabel{emC3}
D(T_{ab}\ast C^{(3)b})=F^{(2)}_{ab}\wedge C^{(3)b},
\end{equation}
\begin{equation}\plabel{emH4}
H^{(4)}_a=g T_{ab}\ast C^{(3)b}+{1\over 8}\e_{a b_1\dots b_4}F^{(2)b_1 b_2}\wedge
F^{(2)b_3 b_4}
\end{equation}
with $H^{(4)a}\equiv D C^{(3)a}=dC^{(3)a}+g A^{(1)a}_{~~~~b}\wedge C^{(3)b}$,
\begin{equation}\plabel{emF2}
D(T^{-1}_{ab}T^{-1}_{cd}\ast F^{(2)ac})=-2g T^{-1}_{a[b}\ast DT_{d]a}-{1\over 2g}
\e_{a_1\dots a_3 bd}F^{(2)a_1 a_2}\wedge H^{(4)a_3}
\end{equation}
$$
+{3\over 2g}\d^{b_1\dots b_4}_{a_1 a_2 bd}F^{(2)a_1 a_2}\wedge
F^{(2)}_{b_1 b_2}\wedge F^{(2)}_{b_3 b_4}-C^{(3)}_b\wedge
C^{(3)}_{d},
$$
\begin{equation}\plabel{emT}
D(T^{-1}_{ab}\ast DT_{bc})=2g^2
(2T_{ab}T_{bc}-T_{bb}T_{ac})\e_{(7)} +T^{-1}_{ad}T^{-1}_{be}\ast
F^{(2)de}\wedge F^{(2)b}_{~~~~~c}
\end{equation}
$$
+T_{cb}\ast C^{(3)b}\wedge C^{(3)}_{a}-{1\over 5}\d_{ac}[2g^2 (2T_{bd}T_{bd}-2(T_{bb})^2)
\e_{(7)} +T^{-1}_{bd}T^{-1}_{ef}\ast F^{(2)df}\wedge F^{(2)eb}+T_{bd}\ast C^{(3)b}\wedge
C^{(3)d}].
$$
These equations, which are the bosonic part of the field equations
of the seven-dimensional supergravity, will be relevant for our
discussions below.

\section{CFT operators from the M5-brane world volume action}

Now let us calculate the CFT operators by expanding the M5 brane action
in the $AdS_7\times S^4$ background.
To be concrete, we restrict our attention to the CFT operators that
correspond to the SUGRA scalar only. The conformal
dimension of these fields is equal to $\triangle=2$ (see, e.g.,
\cite{minwalla}--\cite{seiberg}). Below
we will, following \cite{tran}, restrict to the subsectors of the scalar
matrix $T_{ab}$. The full case is discussed at the end of this
section.

The subsectors we consider are obtained by setting
the gauge fields $A^{(1)}_{ab}$ and $C^{(3)}_{a}$ to zero. Then,
eqs. \p{emC3} -- \p{emT} reduces to
\begin{equation}\plabel{emF2r}
T^{-1}_{a[b}\ast dT_{d]a}=0,
\end{equation}
\begin{equation}\plabel{emTr}
d(T^{-1}_{ab}\ast dT_{bc})=2g^2
[(2T_{ab}T_{bc}-T_{bb}T_{ac})-{1\over 5}\d_{ac}
(2T_{bd}T_{bd}-2(T_{bb})^2)]\e_{(7)}.
\end{equation}
Therefore, this
setting allows one to choose a diagonal parameterization \cite{pope}
for the matrix $T_{ab}$:
\begin{equation}\plabel{Trep}
T_{ab}=diag~ (X_1,\dots,X_5),\qquad \prod_{a=1}^{5} X_a =1
\end{equation}
and
\begin{equation}\plabel{Xrep}
X_a=\exp \left( -{1\over 2}\vec{b}_a \cdot \vec{\phi} \right).
\end{equation}
Here $\vec{\phi}$ is the vector defining four independent scalars
appearing in the reduction from $M^{11}$ to $AdS_7\times S^4$ and
$\vec{b}_a$ are the weight vectors of the fundamental reps. of
$SL(5,R)$ which have the following properties,
\begin{equation}\plabel{bprop}
\vec{b}_a\cdot \vec{b}_b=8\d_{ab}-{8\over 5},\ \ \
\sum_a~\vec{b}_a=0,\ \ \ \sum_a (\vec{u}\cdot \vec{b}_a)\cdot\vec{b}_a=
8\vec{u}
\end{equation}
for an arbitrary vector $\vec{u}$. \footnote{The explicit
representations for the $\vec{b}_a's$  are as follows:
$$
\vec{b}_1=\left( 2,{2\over \sqrt{3}},{2\over \sqrt{6}},
{\sqrt{2}\over \sqrt{5}}\right),\qquad
\vec{b}_2=\left( -2,{2\over \sqrt{3}},{2\over \sqrt{6}},
{\sqrt{2}\over \sqrt{5}}\right),
$$
$$
\vec{b}_3=\left( 0,-{4\over \sqrt{3}},{2\over \sqrt{6}},
{\sqrt{2}\over \sqrt{5}}\right),\qquad \vec{b}_4=\left(
0,0,-\sqrt{6}, {\sqrt{2}\over \sqrt{5}}\right),
$$
$$
\vec{b}_5=\left( 0,0,0,-{4\sqrt{2}\over \sqrt{5}}\right).
$$
After reconstruction of the Lagrangian and the equations of
motions for the scalar fields, the n-point functions of the CFT
operators can be computed, as discussed in \cite{tran}, by use of
the formulae in \cite{fmmr}, \cite{mv}.}

Substituting the diagonal parameterization \p{Trep} into the
metric ansatz \p{nlansds} and expanding it in linear order of
$\vec{\phi}$, we have
$$
ds^2_{11} \simeq \left(1-{1\over 6}\sum_a (\m^a)^2~\vec{b}_a\cdot
\vec{\phi} \right) ds^2_7
$$
\begin{equation}\plabel{ansexp}
+g^{-2}\left( 1+{1\over 3}\sum_c (\m^c)^2~\vec{b}_c \cdot \vec{\phi}\right)
\sum_a \left(1+{1\over 2}\vec{b}_a\cdot \vec{\phi}\right)(d\m^a)^2.
\end{equation}
To make the $SO(5)$ covariance manifest one can rewrite \p{ansexp} in
a coordinate system of Cartesian type, $(x^i,x^a)$,
\begin{equation}\plabel{cart}
\mu^a={x^a\over r}, \qquad r^2=(x^a)^2.
\end{equation}
Note that $g$ is the inverse radius of the $S^4$, i.e. $g^{-1}=R$.

The space-time metric of BPS p-brane
configurations has the form of (see, e.g., \cite{gt}, \cite{ckktp})
\begin{equation}\plabel{ds}
ds^2_{p-brane}=H^{-{2\over p+1}}(dx^i)^2+H^{2\over D-p-3}(dx^a)^2,
\end{equation}
\begin{equation}\plabel{H}
H=1+\left( {R\over r} \right)^{D-p-3},
\end{equation}
where the coordinates, $x^i$, are the brane coordinates and the
coordinates, $x^a$,
are transverse to the brane with $r^2 \equiv(x^a)^2$.
In the near horizon region $r\ll R$ this metric simplifies to the
geometry of an $AdS_{p+2}\times S^{D-p-2}$
\begin{equation}\plabel{ads}
ds^2=\left( {r\over R}\right)^{2(D-p-3)\over p+1} (dx^i)^2+
\left({R\over r}\right)^2 (dx^a)^2.
\end{equation}
For the M5 case the near-horizon region
is $AdS_7\times S^4$, with the metric given by
\begin{equation}\plabel{ads7}
ds^2=\left( {r\over R}\right) (dx^i)^2+
\left({R\over r}\right)^2 (dx^a)^2.
\end{equation}
Using this background metric, \p{ansexp} can be rewritten as
\begin{equation}\plabel{exp1}
ds^2_{11}=grf\sum_i (dx^i)^2+{1\over g^2 r^2}\sum_{a,b=1}^5 g_{ab}dx^a
dx^b
\end{equation}
with
\begin{equation}\plabel{f}
f=1-{1\over 6r^2}\sum_a (x^a)^2\vec{b}_{a}\cdot \vec{\phi},
\end{equation}
\begin{equation}\plabel{gab}
g_{ab}=\d_{ab}+{1\over 2}\vec{b}_a\cdot \vec{\phi}\d_{ab}-{1\over 2r^2}
\vec{b}_a\cdot \vec{\phi}x_a x_b+{1\over 3r^4}\sum_c (x^c)^2
\vec{b}_c\cdot \vec{\phi}\d_{ab}-{1\over 2r^4}\sum_c (x^c)^2
\vec{b}_c\cdot \vec{\phi} x_a x_b.
\end{equation}
There are additional terms coming from \p{ansexp} which are of
second order in $\vec{\phi}$, therefore neglected.

Finally, we expand the action for the M5 \cite{pst},
\cite{blnpst}, \cite{apps}
\begin{equation}\plabel{M5ac}
S=-\int\,d^6 \xi \left[\sqrt{-\det (\hat{g}_{mn}+i\hat{H}_{mn}^{*})}+
{\sqrt{-\hat{g}}\over 4\sqrt{-\widehat{(\partial a)^2}}}\hat{H}^{*mn}\hat{H}_{mnr}
\partial^r a \right]
\end{equation}
$$
+\int_{{\cal M}^6}\, \hat{A}^{(6)}+{1\over 2}db^{(2)}\wedge \hat{A}^{(3)}
$$
around the background defined by \p{exp1} in the small velocities approximation
\cite{ckktp}. In order to do that we need to find the
explicit forms of the
$\hat{A}^{(6)}$ and $\hat{A}^{(3)}$ gauge fields from the expressions of their field
strengths, \p{nlansF4} and \p{nlansF7}. After some algebra one can derive
the following equations,
\begin{equation}\plabel{A3}
\hat{A}^{(3)}=-{1\over 3!g^3}\e_{\a_1\dots
\a_4}{X_{\b}\d_{\b\a_4}\mu^{\a_4} \over
\mu^{0}\tilde{\triangle}}d\mu^{\a_1}\dots d\mu^{\a_3} -{1\over
3!g^3}\e_{\a_1\dots \a_4}{1\over \mu^0 (1+\mu^0)^2}\mu^{\a_1}
d\mu^{\a_2}\dots d\mu^{\a_4} ,
\end{equation}
\begin{equation}\plabel{A6}
\hat{A}^{(6)}=-{1\over 2g}(X^{-1}_{a}\ast dX_{a}(\mu^{a})^2),
\end{equation}
where we have split the index $a=0,1,\dots,4$ into the set of
$(0,\a)$. Note that as in \cite{tran} these are on-shell results
because they hold only up to the equation of motion, \p{emTr}.
However, the on-shell results are sufficient for our purpose.

The small velocity expansion
\footnote{We have used
$\det M=\exp(Tr \ln M)$, which in turn implies
$$
\det (1+M)^{1/2}=1+{1\over 2}Tr M-{1\over 4}[Tr M^2-{1\over 2}(Tr M)^2]+\dots
$$}
leads to
$$
S\approx -\int\,d^6\xi~-{g^3 r\over 2}\sum_a (x^a)^2 \vec{b}_a\cdot \vec{\phi}
$$
\begin{equation}\plabel{M5exp}
+{1\over 2}\sum_{ab}\left({1\over 2}\vec{b}_a\cdot \vec{\phi}\d_{ab}
-{1\over 2r^2}\vec{b}_a\vec{\phi}x_a x_b-{1\over 3r^2}\sum_c (x^c)^2
\vec{b}_c \cdot \vec{\phi}\d_{ab}+\dots \right)\partial_m x^a \partial^m x^b
+\dots,
\end{equation}
where we have omitted the terms of higher order in $\vec{\phi}$ or
derivatives (of $\vec{\phi}$ and $x^a$ as well).

Several remarks are in order concerning how to obtain \p{M5exp}. The general form of
the expansion is
\begin{equation}\plabel{expgen}
S\approx \int\, d^6\xi~ {\cal L}^{(0,0)}+{\cal L}^{(0,1)}+{\cal
L}^{(1,0)}+{\cal L}^{(1,1)} +\dots
\end{equation}
The superscript index, $(p,q)$, indicates the order of
$\vec{\phi}$ and the number of derivatives acting on them and
$x^a$, respectively. Now we will prove that there are no other
terms of the type $(1,0)$ than those we have already given in
(\ref{M5exp}). To this end, note that the induced metric on the M5
worldvolume, which corresponds to the ansatz \p{nlansds}, has the
following form
\begin{equation}\plabel{indg}
\hat{g}_{mn}=\tilde{\triangle}^{1/3}(g_{mn}+\tilde{\triangle}^{-1}T^{-1}_{ab}D_m \mu^a
D_n \mu^b)
\end{equation}
The action for the M5 also involves the inverse worldvolume
metric $\hat{g}^{mn}$, which can be shown to be
\begin{equation}\plabel{invg}
\hat{g}^{mn}=\tilde{\triangle}^{-1/3}\left( g^{mn}-{\tilde{\triangle}^{-1}
T^{-1}_{ab}D^{m}\mu^a D^n\mu^b \over 1+\tilde{\triangle}^{-1}T^{-1}_{ab}D_m \mu^a
D^m \mu^b}\right).
\end{equation}
Up to the terms of the order $(2,0)$ the equations \p{indg} and
\p{invg} can be written as $\hat{g}_{mn}\approx
\tilde{\triangle}^{1/3}g_{mn}$ and $\hat{g}^{mn}\approx
\tilde{\triangle}^{-1/3} g^{mn}$ respectively. The leading terms
in $H^{(3)}$, which come from the first line of eq. \p{M5ac}, are
given by \cite{bns}
\begin{equation}\plabel{Hexp}
S_{H}\approx \int\, d^6\xi \sqrt{-\hat{g}}\left[{1\over 4!}\hat{H}_{mnp}\hat{H}^{mnp}+
{1\over 8 \widehat{{\partial a}{\partial a}}}\partial_m a (\hat{H}^{mnl}-\hat{H}^{*mnl})
(\hat{H}_{nlp}-\hat{H}^{*}_{nlp})\partial^p a+\dots \right].
\end{equation}
They do not contribute to the $(1,0)$ part because \p{Hexp} has
the ``weight" $\tilde{\triangle}^{0}$ that only contributes to the
$(0,0)$, $(2,0)$ and higher order in $\vec{\phi}$ with or without
derivatives. As for the WZ terms, it is clear, from \p{A6}, that
there is no contribution to the $(1,0)$ type terms from the ${\hat
A}^{(6)}$ part. A straightforward calculation also shows that the
contributions of the second term in the WZ part of the action are
solely to $(0,0)$, $(2,0)$ and higher order terms, which completes
our proof.

For the subsectors given by (\ref{Trep}) and (\ref{Xrep}) we have
achieved the goal because the CFT operator has appeared as the
coefficient of $\vec{\phi}$. The coordinates $x^a$ are transverse
to the M5 worldvolume and they are the ones that are identified
with the scalars $\Phi^a$ of the on-shell (2,0) (ultrashort)
supermultiplet.

For the full sectors one should keep the fields, $A^{(1)}_{ab}$
and $C^{(3)}_{a}$. After finding the complete ansatz for
$\hat{A}_3$ and $\hat{A}_6$ and substituting them into the M5
brane action, one again only keeps the terms linear in
$S_{ab}$.\footnote{$S_{ab}$ appeared in (\ref{Treps}).} Finally
one should set all the supergravity modes to zero after taking the
derivative with respect to $S_{ab}$: it is not difficult to see
that the terms that involve $A^{(1)}_{ab}$ and $C^{(3)}_{a}$ will
not be relevant for the final result. Therefore we deduce from
(\ref{M5exp}) the relevant part of the action through the
following chain of relations
$$
S\approx -\int\,d^6\xi~-{g^3 r\over 2}\sum_a (x^a)^2 \vec{b}_a\cdot \vec{\phi}
=-\int\,d^6\xi~g^3 r\sum_a (\Phi^a)^2\left[1-{1\over 2}\vec{b}_a\cdot
\vec{\phi}-1 \right]
$$
$$
\approx -\int\,d^6\xi~g^3 r \sum_a~ \left[e^{-{1\over 2}\vec{b}_a
\vec{\phi}}(\Phi^a)^2-(\Phi^a)^2\right]=-\int\,d^6\xi~g^3 r
\sum_{ab}\left[~(e^S)_{ab} \Phi^a \Phi^b -(\Phi^a)^2\right],
$$
which implies
\begin{equation}\plabel{corr}
S\approx -\int\,d^6\xi~g^3 r  \sum_{ab}
\left( \Phi^a \Phi^b\right)S_{ab}.
\end{equation}

In the boundary region, $r\rightarrow \infty$, $S_{ab}\propto
r^{-1}$ and therefore the boundary condition can be chosen as
\begin{equation}\plabel{Sbc}
S_{ab~\vert_{b.c.}}={1\over r}S^{0}_{ab}.
\end{equation}
Taking the trace constraint on $S_{ab}$ into account we obtain the
CFT operator,
\begin{equation}\plabel{CFT}
{\cal O}^{ab}=(\Phi^a \Phi^b-{1\over 5}\d^{ab}\Phi^c \Phi_c)+\dots
\end{equation}

\section{Conclusions}

Substituting the non-linear ansatz for the eleven-dimensional metric and
 gauge fields into the M5-brane action and expanding it around an
$AdS_7\times S^4$ background we have obtained the CFT
 operators that correspond
to 14 scalars of N=4 D=7 supergravity. The leading terms of the
operators are in the symmetric
traceless representation of the $SO(5)$ R-symmetry group. Therefore,
 our result is consistent, in the leading order, with
the conjecture based on the
superconformal symmetry and Matrix-like model arguments.

However, the CFT operators have subleading terms as well that
include e.g., the (2,0) CFT scalar fields and their derivatives.
Appearance of such terms has been discussed in \cite{frolov} in
the context of type IIB supergravity on $AdS_5\times S^5$. As
noted in \cite{tran} the subleading terms appearing in the CFT
operator could be viewed as in accordance with claim of
\cite{frolov} that supergravity modes are dual to the ``extended"
chiral primary operators. Or/and there could be some field
redefinitions on the CFT side such as the one discussed in
\cite{gkpr}. The interesting problem, therefore, is to compute the
n-point correlators for scalar supergravity modes propagating on
$AdS_7$ by the use of non-linear reduction ansatz \footnote{Two-
and three-point correlators of the (2,0) CFT primaries have been
computed in \cite{lin} in linear ansatz approximation. An
advantage of using the nonlinear ansatz was discussed in
\cite{nv}.} and to check explicitly this observation.

Another problem one can consider is to extend the results obtained
here to another class of CFT operators that correspond to other
supergravity modes and to compare with the results of
\cite{ferrara} based on the primary superfields considerations.

\vspace{1cm} \noindent {\bf Acknowledgements}. \noindent The
authors are grateful to Ergin Sezgin and Chris Pope for interest in
this work and encouragement. AJN is thankful to Vladimir Akulov, Igor Bandos
and Dmitri Sorokin for fruitful and illuminating discussions. The
work of AJN is supported in part by the NSF Grant PHY-0070964,
by INTAS under a Call 2000 Project N254 and by the Ukrainian
Ministry of Science and Education Grant N2.51.1/52-F5/1795-98.
The work of IYP is supported by US Department of Energy under grant
DE-FG03-95ER40917.

\end{document}